\documentclass[12pt]{article}
\usepackage{graphicx}
\def\ltap{\raisebox{-.4ex}{\rlap{$\sim$}} \raisebox{.4ex}{$<$}}

\def\journal{\topmargin 0in   \oddsidemargin 0in
        \headheight 0pt \headsep 0pt
        \textwidth 6.5in 
\textheight 9in 
        \marginparwidth 1.5in
        \parindent 2em
        \parskip .5ex plus .1ex         \jot = 1.5ex}
\journal

\def\ra{\rightarrow}
\begin{document}
\begin{titlepage}
\begin{center}
October 17, 2000      \hfill    LBNL-45962\
\vskip .5in

{\large \bf New Physics and the Landau Pole}
\footnote
{This work is supported in part by the Director, Office of Science, Office
of High Energy and Nuclear Physics, Division of High Energy Physics, of the
U.S. Department of Energy under Contract DE-AC03-76SF00098}

\vskip .5in

Michael S. Chanowitz\footnote{Email: chanowitz@lbl.gov}

\vskip .2in

{\em Theoretical Physics Group\\
     Ernest Orlando Lawrence Berkeley National Laboratory\\
     University of California\\
     Berkeley, California 94720}
\end{center}

\vskip .25in

\begin{abstract}

In scalar field theories the Landau pole is an ultraviolet singularity 
in the running coupling constant that indicates a mass scale at which 
the theory breaks down and new physics must intervene.  However, 
new physics at the pole will in general affect the running of the low 
energy coupling constant, which will in turn affect the location of the 
pole and the related upper limit (``triviality'' bound) on the low 
energy coupling constant.  If the new physics is strongly coupled to 
the scalar fields these effects can be significant even though they 
are power suppressed.  We explore the possible range of such effects 
by deriving the one loop renormalization group equations for an 
effective scalar field theory with a dimension 6 operator representing 
the low energy effects of the new physics.  As an independent check we 
also consider a renormalizable model of the high-scale physics 
constructed so that its low energy limit coincides with the effective 
theory.

\end{abstract}

\end{titlepage}

\renewcommand{\thepage}{\roman{page}}
\setcounter{page}{2}
\mbox{ }

\vskip 1in

\begin{center}
{\bf Disclaimer}
\end{center}

\vskip .2in

\begin{scriptsize}
\begin{quotation}
This document was prepared as an account of work sponsored by the United
States Government. While this document is believed to contain correct
 information, neither the United States Government nor any agency
thereof, nor The Regents of the University of California, nor any of their
employees, makes any warranty, express or implied, or assumes any legal
liability or responsibility for the accuracy, completeness, or usefulness
of any information, apparatus, product, or process disclosed, or represents
that its use would not infringe privately owned rights.  Reference herein
to any specific commercial products process, or service by its trade name,
trademark, manufacturer, or otherwise, does not necessarily constitute or
imply its endorsement, recommendation, or favoring by the United States
Government or any agency thereof, or The Regents of the University of
California.  The views and opinions of authors expressed herein do not
necessarily state or reflect those of the United States Government or any
agency thereof, or The Regents of the University of California.
\end{quotation}
\end{scriptsize}

\vskip 2in

\begin{center}
\begin{small}
{\it Lawrence Berkeley National Laboratory is an equal opportunity employer.}
\end{small}
\end{center}

\newpage

\renewcommand{\thepage}{\arabic{page}}
\setcounter{page}{1}
\noindent {\bf (1) Introduction }

In a classic paper Dashen and Neuberger\cite{dn} showed in 
perturbation theory at one loop that the location of the Landau pole 
in scalar field theory implies an upper limit on the mass of the Higgs 
boson.  The Landau pole indicates the mass scale at which the running 
coupling constant, $\lambda_{Q}$, diverges.  In the elegantly simple 
DN (Dashen-Neuberger) analysis it implies an upper bound on the scale 
where new physics supplants the scalar field theory, which is regarded 
as an effective low energy description of the Higgs sector.  By 
requiring a minimal hierarchy between the new physics scale $\Lambda$ 
and the Higgs boson mass, e.g., $\Lambda \ge 2m_{H}$, DN obtained an 
upper bound on $m_{H}$ from the perturbative relation between the low 
energy coupling constant $\lambda$ and the ratio $\Lambda/m_{H}$.  
They proposed a space-time lattice ``experiment'' to confirm the bound 
and make it quantitative.  Lattice calculations\cite{lattice} have 
established the bound on $m_{H}$ at about 700 GeV, not far from the 
$\simeq 1$ TeV estimate of DN.

The purpose of this paper is to explore, in a similarly transparent
way using one loop perturbation theory, the extent to which the new
physics that must occur at or below the Landau pole can affect the
relationship between the pole location and the low energy coupling
constant.  In this paper we consider the simplest case: ${\rm O}(N)\
\phi^{4}$ field theory in the symmetric phase, for which the DN
analysis implies an upper bound on the coupling constant.  The broken
symmetry phase will be considered elsewhere.

Since new physics must exist at the Landau pole, it is not optional
but essential to consider its possible effect on the analysis.  The
obvious method is to introduce higher dimension operators to represent
the power-suppressed, low energy effects of the new physics.  For
instance, effects of dimension 6 operators are suppressed by
$\mu^{2}/\Lambda_{\rm Landau}^{2}$ where $\mu$ is the low energy
renormalization scale and $\Lambda_{\rm Landau}$ is the scale of the
Landau pole.\footnote {Away from the triviality limit the new physics
could lie below the pole, $M_{\rm New} < \Lambda_{\rm Landau}$, in
which case the effects of the new physics would be larger, $\propto
\mu^{2}/M_{\rm New}^{2}$.}
For the minimal hierarchy used to obtain the upper bound this 
suppression is only a factor 1/4, which could be overcome if the new 
physics is strongly coupled to the low energy scalar sector.  We will 
compute the effect of such an operator on the running of the scalar 
coupling constant and the position of the Landau singularity.

Most lattice studies of triviality (e.g., \cite{lattice}) considered
renormalizable scalar field theories without higher dimension
operators representing the possible effects of new physics and would
apply literally if the new physics at the pole were actually the
assumed space-time lattice.  Some lattice simulations\cite{hknv} (of
the Higgs phase) have explored the effects of new physics by
introducing higher dimension operators, as we do here, but with a
different focus.  Their results agree qualitatively with ours but are
not directly comparable for two reasons (in addition to the fact that
different phases are considered).  First, a precise comparison would
require studying the same operators with carefully matched
normalizations.  Second, the goal in \cite{hknv} is to establish an
upper limit on $m_{H}$ such that corrections to various Higgs sector
quantities (e.g., the Higgs decay width) from the higher dimension
operators are limited to a few percent, whereas the focus in this
paper is on the maximum allowed value, independent of the size of
other corrections, which are not known experimentally and could
actually be large.

The coupling constants must be defined by renormalization conditions.
We define the scalar coupling constant $\lambda$ in terms of the $2
\ra 2$ scattering amplitude, for off-shell external momenta, as is
customary in the RG (renormalization group) analysis in order to avoid
mass singularities.\cite{swtext} Then the first dimension 6 operator
that comes to mind, $\propto \phi^{6}/ \Lambda^{2}$, does not
contribute to the running of $\lambda_{Q}$, since its one loop
contribution to the scattering amplitude, shown in figure 1, is a
(divergent) constant, independent of the external scale $Q$.  For the
off-shell renormalization condition adopted below there is just one
other independent O(N) symmetric dimension 6 operator, which we choose
to write in the form $\kappa (\partial ({\bf \phi})^{2})^{2}$.  Here
$\kappa=C /\Lambda^{2}$ where $C$ is a dimensionless constant and
$\Lambda$ is the mass scale of the high energy theory, which we
identify with the position of the Landau pole.  Using the off-shell
renormalization condition, $\kappa$ is also defined in terms of the $2
\ra 2$ amplitude.  Operator mixing occurs, resulting in coupled
renormalization group equations for $\lambda$ and $\kappa$ which we
compute to order $\lambda^{2}$ and $\lambda\kappa$.  Solving the
coupled RGE's (renormalization group equations) we find fractional
corrections of order $\kappa\mu^{2}/\lambda$ to the Landau pole
postion, $\Lambda_{\rm Landau}$, and to the upper limit on the low
energy coupling $\lambda$, where $\mu$ is the low energy
renormalization scale, chosen to be the scalar mass.

While effective Lagrangians were first used strictly in tree 
approximation, it has long been realized that it makes sense to 
consider them at the quantum level.\cite{sw1} Though technically 
``nonrenormalizable'' in the sense that they cannot be defined to all 
orders by a finite number of renormalization conditions, they can be 
renormalized to any finite order.  The quantum effects of chiral 
effective Lagrangians have been thoroughly analyzed at the one loop 
level\cite{gl} and one loop quantum corrections from dimension 6 
operators have been used to study the possible consequences of new 
physics in electroweak gauge theories.\cite{eweff} These calculations 
can be carried out to useful approximations, though arbitrary 
precision would require arbitarily many renormalization 
constants.  This is not a concern, since arbitrary precision is in any 
case not the goal in applications of effective theories.  See 
\cite{am} for an excellent review with interesting examples and 
additional references.

We have verified the renormalization of the effective theory 
considered here, first by checking explicitly that the result is 
independent of the choice of regulator (for dimensional 
regularization, Pauli-Villars regularization, or Euclidean space 
cutoff) and second by obtaining the same result from a renormalizable 
model with an additional, heavy O(N) singlet scalar field, 
constructed so that its low energy limit corresponds to the effective 
theory.  Because of the dimension 6 operator the effective theory has 
quadratic and logarithmic divergences at one loop.  The quadratic 
divergences are constants, independent of the renormalization scale 
and are absorbed into the $\delta \lambda$ counterterm without 
affecting the running of $\lambda$.  Furthermore, as the 
renormalizable model makes clear, the quadratic divergences are in any 
case artifacts of the effective theory, dominated by the scale of the 
cutoff where the effective theory breaks down.  In contrast the 
logarithmic divergences reflect the domain in which the effective 
theory is valid and may be reliably extracted from the effective 
theory.  They give rise to the renormalization scale dependence from 
which the RGE's follow.

Section 2 presents a brief review of the DN analysis, modified 
slightly to apply to the symmetric phase.  In section 3 we derive the 
one loop coupled RGE's for the effective theory.  In section 4 the 
results are rederived from the renormalizable model.  The coupled 
RGE's are solved in section 5.  In section 6 we use the solutions to 
estimate the corrections to the Landau pole position and to the 
triviality bound in the strong coupling regime.  We conclude with a 
brief discussion in section 7.  

\noindent {\bf (2) The DN analysis }

We review the DN analysis, considering both the broken symmetry phase 
of O(4) scalar field theory (the SM Higgs sector) as considered by DN 
and also the unbroken phase which is the focus of this paper.  For 
renormalizable $\phi^{4}$ theory the ultraviolet RG behavior of the 
two phases is the same and the DN analysis applies also to the 
symmetric phase.  However we must modify the renormalization 
conditions slightly, since DN used the Higgs boson mass $m_{H}$ to 
specify the low energy coupling $\lambda$.  In order to have a method 
that applies also to the symmetric phase we will define the low energy 
coupling constant in terms of the $2 \ra 2$ scattering amplitude.

Where $\phi$ is an N component scalar field the Lagrangian is
$$
{\cal L} = {1 \over 2}(\partial {\phi})^{2} 
          - {\lambda \over 4} ( {\phi}^{2})^{2} 
          - {\mu^{2} \over 2}{\phi}^{2}.    \eqno{(2.1)} 
$$
We first consider the symmetric phase, $\mu^{2}>0$.  At the quantum 
level, with the RG analysis in mind, we define the renormalized 
coupling constant, $\lambda = \lambda_{\mu}$, by an off-shell 
scattering amplitude\cite{swtext} chosen for convenience at a 
symmetric, space-like scale
$$
s=t=u=-\mu^{2},   \eqno{(2.2)}  
$$
that is, 
$$
{\cal M}(\phi_{1} \phi_{1} \ra \phi_{1} \phi_{1})_{s=t=u=-\mu^{2}}=
        - 6i\lambda.                  \eqno{(2.3)} 
$$
The one loop amplitude (see figure 2a) at an arbitrary space-like scale 
$Q^{2}<0$ is
$$
{\cal M}_{s=t=u=Q^{2}}^{(1)}= 
  3\left( (-6i\lambda)^{2} + (N-1)(-2i\lambda)^{2}\right )
     {i^{2}\over 2}
     \int{d^{4}k\over (2\pi)^{4}} {1\over (k^{2}-\mu^{2})
                                               ((k+Q)^{2}-\mu^{2})} 
                               \eqno{(2.4)}
$$
Rewriting the integrand as a parametric integration 
$\int dx$ and performng the $k$ integration by dimensional 
regularization in $n$ dimensions the integral in (2.4) becomes
$$
{i\over 16\pi^{2}}\Gamma(\epsilon) \int^{1}_{0}dx 
                          \left( {(x^{2}-x)Q^{2} - 
                          \mu^{2} \over \mu_{D}^{2}}\right)^{-\epsilon}
              \eqno{(2.5)}
$$
where $\Gamma$ is the Gamma function, $\epsilon = 2 - {n\over 2}$ and
$\mu_{D}$ is the regularization scale.  The $x$ integration is then
evaluated for $-Q^{2}\gg \mu^{2}$, with the result
$$
{\cal M}_{Q}^{(1)}= {3i(N+8)\lambda^{2} \over 8\pi^{2}}
     \left(\Gamma(\epsilon) - {\rm log}{Q^{2}\over \mu_{D}^{2}}\right)
     + \ldots       \eqno{(2.6)}
$$
where the omitted terms in (2.6) are finite constants that will be 
absorbed by counterterms without affecting the running of 
$\lambda_{Q}$.

Using the method of ``renormalized perturbation theory'', 
\cite{peskin} we introduce counterterms,
$$
{\cal L}_{\rm CT}= 
        {\delta Z \over 2}(\partial \bf {\phi})^{2} 
          - {\delta \lambda \over 4} (\bf {\phi}^{2})^{2} 
          - {\delta \mu^{2} \over 2}\bf {\phi}^{2},    \eqno{(2.7)}
$$          
so that the amplitude through one loop is 
$$
{\cal M}_{Q}= - 6i(\lambda + \delta \lambda ) 
                                  + {\cal M}_{Q}^{(1)}.   \eqno{(2.8)}
$$
The counterterm $\delta \lambda$ is then determined from the definition 
of $\lambda$, equation (2.3), to be 
$$
\delta \lambda= -{i\over 6}{\cal M}_{\mu}^{(1)}.  \eqno{(2.9)}
$$
Defining the running coupling constant at scale $Q$ as
$$
\lambda_{Q}  = {i \over 6}{\cal M}_{Q}          \eqno{(2.10)}
$$
we then find 
$$
\lambda_{Q}= \lambda  
              +{i\over 6}\left( {\cal M}_{Q} 
              - {\cal M}_{\mu}\right)              
              = \lambda  
                 + {N+8 \over 8\pi^{2}}\lambda^{2}{\rm log}{Q\over \mu}.
                              \eqno{(2.11)}
$$

The wave function and mass renormalizations can be neglected because 
they are trivial in the $\phi^{4}$ model at one loop: the wave 
function renormalization vanishes and the mass is renormalized by a 
$Q$ independent constant that is absorbed in the mass counterterm.

From (2.11) it is easy to determine the Landau pole and the upper bound 
on $\lambda = \lambda_{\mu}$. Differentiating (2.11) we have the RGE
$$
{d\lambda_{Q}\over d{\rm log}Q} = b_{N}\lambda^{2} 
           =b_{N}\lambda^{2}_{Q} + {\rm O}(\lambda^{3})    \eqno{(2.12)}
$$
where 
$$
b_{N}= {N+8 \over 8\pi^{2}}.         \eqno{(2.13)}
$$
Integrating (2.12) from $\mu$ to $Q$ we have
$$
\lambda_{Q}= { \lambda \over 1-b_{N}\lambda \  
                   {\rm log}\left({Q\over \mu}\right)  }
                   \eqno{(2.14)}
$$
which exhibits the pole at 
$$
{\rm log}\left({\Lambda_{\rm Landau}\over \mu}\right) = 
                  {1\over b_{N}\lambda}
                   \eqno{(2.15)}
$$
The upper bound on the coupling constant then follows by requiring a 
minimal hierarchy between $\Lambda_{\rm Landau}$ and $\mu$. 
For instance, 
$$
\Lambda_{\rm Landau} > 2\mu   \eqno{(2.16)}
$$ 
implies 
$$
\lambda < {1\over b_{N}\  {\rm log}2}.     \eqno{(2.17)}
$$ 

In the broken symmetry phase, $\mu^{2}<0$, the analysis proceeds 
as above with the low energy renormalization specified at 
$-m_{H}^{2}$ instead of $-\mu^{2}$.\footnote
{Since we have neglected 
$\mu^{2}<<Q^{2}$ as noted above, the fact that we now have 
Goldstone boson loops in addition to the Higgs boson loop has no 
effect on the quoted results.}
Since the Higgs boson mass is proportional to the coupling,
$$
m_{H}^{2}= 2\lambda v^{2},     \eqno{(2.18)}
$$ 
where $v^{2}= 4m_{W}^{2}/g^{2}$ is determined from the $W$ boson mass 
and SU(2) gauge coupling constant, the upper bound on $\lambda$ becomes 
an upper bound on $m_{H}$.  Setting $b_{N}=3/2\pi^{2}$ for N=4 we obtain 
the DN bound,
$$ 
    m_{H}^{2} < { 4\pi^{2}v^{2} \over 3\ {\rm log} {\Lambda_{\rm
          Landau}\over m_{H}}  }
\eqno{(2.19)}
$$ 
For $\Lambda_{\rm Landau} > 2m_{H}$ this implies $m_{H}\ \ltap\ 1.08$ 
TeV.

\noindent {\bf (3) The effective theory }

The effective theory is defined by
$$
{\cal L}_{\rm EFF} = {1 \over 2}(\partial { {\phi}})^{2} 
          - {\lambda^{E} \over 4} ({ {\phi}}^{2})^{2} 
          - {\mu^{2} \over 2}{ {\phi}}^{2}
          + {\kappa \over 4} (\partial { {\phi}}^{2})^{2}. 
                                                         \eqno{(3.1)} 
$$
where ${ \phi}$ is an $N$ component scalar field, and the superscript 
$E$, for ``effective,'' distinguishes $\lambda^{E}$ from the coupling 
$\lambda^{R}$ of the renormalizable theory defined in the next 
section.  The coupling $\kappa$ is dimensionful, $\kappa = C/M^{2}$, 
where $M$ is the mass scale of the ``new physics'' that gives rise to 
the dimension 6 operator and $C$ is a dimensionless constant 
characterizing the strength of the interaction between the new physics 
and the scalar sector. 

There is another independent dimension 6 operator that is quadratic in
momentum, which may be written as $\phi^2 (\partial \phi)^2$.  On
mass-shell it can be expressed as a linear combination of the
dimension 6 operator in equation (3.1) plus the $\phi^{4}$
interaction.  Off-shell it is in general an independent operator.
However for the symmetric off-shell renormalization condition
specified below in equation (3.3), its contribution is proportional to
the dimension 6 operator in (3.1) and it is not considered separately
in our analysis.

We define the renormalized couplings in terms of the diagonal elastic 
scattering amplitude ${\cal M}(\phi_{1} \phi_{1} \ra \phi_{1} 
\phi_{1})$ so that the definition can be used for all $N\geq 1$.  The 
tree approximation amplitude from ${\cal L}_{\rm EFF}$ is
$$
{\cal M}(\phi_{1} \phi_{1} \ra \phi_{1} \phi_{1})=
        - 6i\lambda^{E}   + 2i\kappa(s+t+u).               
        \eqno{(3.2)} 
$$
Since $s+t+u=\Sigma p_{i}^{2}=4\mu^{2}$ for on-shell scattering, the 
on-shell amplitude is indistinguishable from the amplitude due to the 
$\lambda\phi^{4}$ interaction alone with $\lambda$ replaced by 
$\lambda ^{E} - {4\over 3}\kappa \mu^{2}$.  The $(\partial { 
{\phi}}^{2})^{2}$ and $\phi^{4}$ interactions can however be 
distinguished by other means, for instance, with the off-shell 
four-point function or the on-shell six-point function.  Since we wish 
in any case to consider an off-shell configuration to avoid mass 
singularities in the RG analysis\cite{swtext}, we will use the 
off-shell four-point function to define $\kappa$ and $\lambda$.

In this section we will renormalize the effective Lagrangian at one
loop order and to leading order in $\kappa$, retaining terms of order
$\lambda^{2}$ and $\lambda\kappa$.  We compute the running coupling
constants $\lambda_{Q}^{E}$ and $\kappa_{Q}$, where $Q$ is the
renormalization scale defined below.  Wave function and mass
renormalization can be ignored to this order, since both contribute
constants, independent of $Q$.  In the renormalizable $O(N)$
$\phi^{4}$ field theory in the symmetric phase, considered in section
2 above, the wave function renormalization vanishes and the mass
renormalization is accomplished by just a $Q$ independent counterterm.
With the addition of the dimension 6 operator in equation (3.1), the
wave function renormalization does not vanish but is constant so that,
as in the renormalizable $\phi^{4}$ theory, no anomalous dimension is
induced for the field $\phi$.  The mass renormalization in ${\cal
L}_{\rm EFF}$ also involves only a $Q$ independent counterterm.
Neither has any effect on the running of the coupling constants and so
can be ignored.  These conclusions follow because the one loop
integral represented by figure 3 has no
dependence on the external scale other than the multiplicative factor
of $Q^{2}$ from the vertex of the dimension 6 operator.

To specify the renormalization conditions for $\lambda^{E}$ and 
$\kappa$ it is convenient to choose a symmetric, spacelike, off-shell 
point
$$
s = t = u = Q^{2} < 0,        \eqno{(3.3)}
$$ 
corresponding to spacelike external 4-momenta with $p_{i}^{2}= {3\over 
4}Q^{2}$ for each external leg, $i=1,2,3,4$.  Then the low energy 
coupling constants $\lambda^{E} = \lambda_{\mu}^{E}$ and 
$\kappa=\kappa_{\mu}$ are defined by
$$
{\cal M}(\phi_{1} \phi_{1} \ra \phi_{1} \phi_{1})_{Q^{2}=-\mu^{2}}=
        - 6i(\lambda ^{E}   + \kappa \mu^{2}),            
        \eqno{(3.4)} 
$$
and the renormalized running couplings $\lambda _{Q} ^{E}$ and $\kappa 
_{Q}$ for arbitrary $Q$ are defined by 
$$
{\cal M}(\phi_{1} \phi_{1} \ra \phi_{1} \phi_{1})_{Q}=
        - 6i(\lambda _{Q} ^{E}   - \kappa _{Q} Q^{2}).            
        \eqno{(3.5)} 
$$
Actually we must vary $Q^{2}$ by a small amount around each given 
central value since at least two measurements are needed to determine 
both $\lambda^{E}_{Q}$ and $\kappa_{Q}$, 
e.g., $Q^{2}= Q^{2}_{\rm Central} \pm \epsilon$ where $\epsilon \ll 
 Q^{2}_{\rm Central}$.  This is a difficult task, 
but our excellent, highly paid gedanken experimenters have the 
necessary skills to carry out the measurements (e.g., using dispersion 
relations in the external off-shell masses).

We now compute the one loop renormalized couplings to order
${\lambda^{E}}^{2}$ and $\lambda^{E}\kappa$.  The relevant Feynman
diagrams are shown in figure 2.  We will regularize the loop integrals
with a momentum space cutoff, since it provides the most physical
description of the loop amplitudes in the effective theory.  We have
the luxury of this choice since we are not concerned here with gauge
invariance.  We have checked that the same results are obtained from
dimensional and Pauli-Villars regularization.

Since regularization by cutoff is becoming a lost art (see however 
\cite{peskin}), we will warm up by evaluating the three ${{\rm 
O}(\lambda^{E}}^{2})$ diagrams, which were computed by dimensional 
regularization in section 2.  Together they contribute
$$
\delta {\cal M}_{Q}^{(\lambda^{2})}= 
  6(N+8){\lambda^{E}}^{2}
     \int{d^{4}k\over (2\pi)^{4}} {1\over (k^{2}-\mu^{2})
                                               ((k+Q)^{2}-\mu^{2})} 
                               \eqno{(3.6)}
$$
or, introducing the parametric integral over $x$,
$$
\delta {\cal M}_{Q}^{(\lambda^{2})} = 6 (N+8){\lambda^E}^{2}
                            \int^{1}_{0}dx \int{d^{4}k\over (2\pi)^{4}}
          {1   \over       (k^{2} - X)^{2} }
                               \eqno{(3.7)}
$$          
where 
$$
X= (x^{2}-x)Q^{2} + \mu^{2}
                               \eqno{(3.8)}
$$ 

To define the integral with a cutoff we first rewrite it as a
Euclidean space integral by continuing the integrand into the 
complex $k_{0}$ plane. By contour integration we relate the integral 
along the real $k_{0}$ axis to one along the imaginary axis:
$$
\int^{\infty}_{-\infty}dk_{0} f(k_{0}) = 
-\int^{-i\infty}_{i\infty}dk_{0} f(k_{0}) = 
-i \int^{\infty}_{-\infty}dk_{0}^{\prime} f(-ik_{0}^{\prime}),      
                               \eqno{(3.9)}
$$
where we define $k_{0}^{\prime} = -ik_{0}$.  The arcs in the first and 
third quadrants may be neglected since they only contribute constants 
that are absorbed in the counterterms.  The Minkowski space 4-vector 
$k$ then becomes negative definite within the domain of the 
$k_{0}^{\prime}$ integration,
$$
k^{2}= -{k_{0}^{\prime}}^{2} -\vec k^{2}.  \eqno{(3.10)}
$$ 

We define a Euclidean 4-vector $k_{E}$ whose components are 
$$
k_{E}=  (k_{0}^{\prime}, \vec k)  \eqno{(3.11)}
$$ 
so that 
$$
k^{2}= -k_E^{2} .  \eqno{(3.12)}
$$ 
The Minkowski space integral, equation (3.7), is then replaced by a 
Euclidean space integral,
$$
\delta {\cal M}_{Q}^{(\lambda^{2})}=  -6i(N+8){\lambda^E}^{2}
     \int^{1}_{0}dx \int{d^{4}k_{E}\over (2\pi)^{4}}
          {1   \over       (k^{2}_{E} + X)^{2} } 
                               \eqno{(3.13)}
$$
which we will regulate with the O(4) symmetric cutoff $k_{E}^{2} \leq 
\Lambda^{2}$.  Equation (3.13) exhibits the advantage of choosing 
off-shell, spacelike external momenta: for $Q^{2} < 0$ we have $X > 0$ 
so that $k^{2}_{E} + X$ never vanishes and the integrand has no 
singularities.

The integrand is spherically symmetric so the angular 
integration yields a factor $2\pi^{2}$ and the remaining integration 
over $|k^{2}_{E}| \leq \Lambda^{2}$ is easily done.  The result is
$$
\delta {\cal M}_{Q}^{(\lambda^{2})}=  
     {-3i{\lambda^E}^{2}(N+8)\over 8\pi^{2}}
     \int^{1}_{0}dx {\rm log}\left( {X \over \Lambda^{2}}\right)
                               \eqno{(3.14)}
$$
where we omit terms of order $1/\Lambda^{2}$.  Finally, for $Q^{2}\gg 
\mu^{2}$, we approximate ${\rm log}(X) \simeq {\rm log}(Q^{2})$, 
obtaining the usual result,
$$
\delta {\cal M}_{Q}^{(\lambda^{2})}=  
     {-3i{\lambda^E}^{2}(N+8)\over 4\pi^{2}}
     {\rm log}\left( {Q \over \Lambda}\right)
                               \eqno{(3.15)}
$$
where $Q = \sqrt{-Q^{2}}$.  The terms we have neglected by 
approximating ${\rm log}(X) \simeq {\rm log}(Q^{2})$ are either 
constants that would be absorbed in counterterms or are suppressed by 
$\mu^{2}/Q^{2}$.  The logarithmic term in (3.15) agrees with the 
dimensional regularization result (2.6) if the cutoff $\Lambda$ is 
identified with the dimensional regularization scale $\mu_{D}$.

It is straightforward to apply the same method to the O($\lambda
\kappa$) diagrams shown in figure 2b, of which there are six, each
contributing equally due to the symmetric kinematics, equation (3.3).
Including a factor 6 for the number of diagrams, the Feynman rules
yield
$$
\delta {\cal M}_{Q}^{(\lambda \kappa)}=
     -36 \lambda^{E} \kappa
       \int{d^{4}k\over (2\pi)^{4}}
       {(k+p_{1})^{2} +  (k+p_{2})^{2} +{1\over 3}(N+2)Q^{2}
       \over (k^{2}-\mu^{2}) ((k+Q)^{2}-\mu^{2})} 
                               \eqno{(3.16)}
$$
Introducing the Feynman parameter integration and symmetrizing the $k$ 
integration, we have
$$
\delta {\cal M}_{Q}^{(\lambda \kappa)}=
     -36 \lambda ^{E} \kappa
        \int^{1}_{0}dx \int{d^{4}k\over (2\pi)^{4}}
       {2k^{2} + (2x^{2} -2x + {N\over 3} +{13\over 6}) Q^{2}
       \over    (k^{2} - X)^{2} } 
                               \eqno{(3.17)}
$$
where $X$ is defined in equation (3.8). Just as in equations (3.9 - 
3.13), the $k$ integration is expressed as a Euclidean space 
integral, which after the angular integration is 
$$
\delta {\cal M}_{Q}^{(\lambda \kappa)}=
     { -9i\lambda ^{E} \kappa \over 4\pi^{2}}
        \int^{1}_{0}dx 
        \int^{\Lambda^{2}}_{0} dk_{E}^{2} k^{2}_{E}
       {-2k^{2}_{E} +(2x^{2} -2x + {N\over 3} +{13\over 6}) Q^{2}
       \over    (k^{2}_{E} + X)^{2} } 
                               \eqno{(3.18)}
$$

The term proportional to $Q^{2}$ diverges logarithmically.  The term 
proportional to $k^{2}$ is rewritten to isolate the quadratic 
divergence,
$$
\int^{\Lambda^{2}}_{0} dk_{E}^{2} 
     { (k^{2}_{E})^{2} \over    (k^{2}_{E} + X)^{2} } = 
     \Lambda^{2} - \int^{\Lambda^{2}}_{0} dk_{E}^{2} 
     { 2k_{E}^{2}X  +X^{2}\over    (k^{2}_{E} + X)^{2} }. 
                               \eqno{(3.19)}
$$
The remaining log divergent term in (3.19) is then combined with the 
log divergent term proportional to $Q^{2}$ in (3.18). The 
finite term proportional to $X^{2}$ is neglected since it contributes 
O($\Lambda^{-2}$) to the amplitude. The result is 
$$
\delta {\cal M}_{Q}^{(\lambda \kappa)} =
     { -9i\lambda ^{E} \kappa \over 4\pi^{2}}\left( -\Lambda^{2}
     +  \int^{1}_{0}dx 
        \int^{\Lambda^{2}}_{0} dk_{E}^{2} k^{2}_{E}
       { (6x^{2} -6x + {N\over 3} +{13\over 6}) Q^{2} +4\mu^{2}
       \over    (k^{2}_{E} + X)^{2} } \right)
                                   \eqno{(3.20)}
$$
Performing the integral over $k_{E}^{2}$ we are left with
$$
\delta {\cal M}_{Q}^{(\lambda \kappa)} =
     { 9i\lambda ^{E} \kappa \over 4\pi^{2}}\left( \Lambda^{2}
     +  \int^{1}_{0}dx {\rm log}\left ({X\over \Lambda^{2}} \right)
      \left \{(6x^{2} -6x + {N\over 3} +{13\over 6}) Q^{2} +4\mu^{2}
       \right\} \right)  
                                   \eqno{(3.21)}
$$

If the cutoff $\Lambda$ is chosen to be equal to the scale of the new 
physics, $\Lambda \simeq M$, then the $\Lambda^{2}$ term in (3.21) is 
an artifact since it is dominated by the region of the $k$ integration 
near $M$ where the effective theory fails.  In any case, since it is 
independent of $Q$, it does not affect the running of the couplings.  
By contrast the log($\Lambda$) term samples the entire hierarchy 
between $M$ and $Q$: e.g., for large log($\Lambda/Q$) the region 
between $\Lambda$ and $\Lambda/2$ only contributes a small fraction, 
log$(2) \ll {\rm log}(\Lambda/Q)$ to the logarithm while contributing 
3/4 of the quadratic term.  This is seen again in the renormalizable 
model presented in the next section.

The running couplings $\lambda^{E}_{Q}$ and $\kappa_{Q}$ receive 
contributions from the terms in (3.21) proportional to 
$\mu^{2}$log($Q$) and $Q^{2}$log($Q$) respectively.  However, the 
contribution proportional to $\mu^{2}$log($Q$) is not given simply by 
the coefficient of the term proportional to $\mu^{2}$ in (3.21), since 
an additional $\mu^{2}$ contribution is hidden in the term 
proportional to $Q^{2}$.  Evaluating the integrals over $x$ through 
order $\mu^{2}/Q^{2}$, we have
$$
\int^{1}_{0}dx\ {\rm log}\left ({X\over \Lambda^{2}} \right) = 
     \left(1 - 2{\mu^{2} \over Q^{2}} \right)
     {\rm log}\left({Q^{2}\over \Lambda^{2}}\right)  
     + {\rm O}\left({\mu^{2}\over Q^{2}}\right)^{2} +\ldots      
                                                                \eqno{(3.22)}
$$
and for integer $n\ge 1$ 
$$
\int^{1}_{0}dx\ x^{n}{\rm log}\left ({X\over \Lambda^{2}} \right) = 
       {1\over n+1}\left(1 - (n+1){\mu^{2} \over Q^{2}} \right)
       {\rm log}\left({Q^{2}\over \Lambda^{2}}\right) 
       +{\rm O}\left({\mu^{2}\over Q^{2}}\right)^{2}+\ldots      
                                                                \eqno{(3.23)}
$$
where we also omit constant terms that are independent of log($Q$)
and do not affect the running of $\lambda$ and $\kappa$.
Substituting (3.22 - 3.23) into (3.21) we have
$$
\delta {\cal M}_{Q}^{(\lambda \kappa)} =
     { -3i\lambda ^{E} \kappa \over 4\pi^{2}}
     {\rm log}\left ({Q\over \Lambda} \right)
      \left \{ 2(2N +1)\mu^{2} 
              - (2N +7)Q^{2} \right\} +\ldots
                                   \eqno{(3.24)}
$$
where $Q=|Q|=\sqrt{-Q^{2}}$ and we again omit all terms 
(including the $\Lambda^{2}$ term) that do not contribute to the 
running of $\lambda$ and $\kappa$.

We can now derive the renormalized couplings and the RGE's.  The 
counterterm Lagrangian for the effective theory is
$$
{\cal L}_{\rm CT}= 
        {\delta Z \over 2}(\partial {\phi})^{2} 
                   +{\delta \kappa \over 4} (\partial  {\phi}^{2})^{2}  
 - {\delta \lambda^{E} \over 4} ( {\phi}^{2})^{2}
          - {\delta \mu^{2} \over 2} {\phi}^{2},    \eqno{(3.25)}
$$ 
As noted above, $\delta Z$ and $\delta \mu$ have no $Q$ dependence and 
can be ignored. Combining the relevant tree, counterterm, and one loop 
contributions, the amplitude is 
$$
{\cal M}_{Q}= -6i(\lambda^{E} + \delta \lambda^{E})
                        +6i(\kappa + \delta \kappa )Q^{2}
                        +\delta {\cal M}_{Q}^{(1)}
                                                 \eqno{(3.26)} 
$$
where
$$
   \delta {\cal M}_{Q}^{(1)} = \delta {\cal M}_{Q}^{(\lambda^{2})}
   + \delta {\cal M}_{Q}^{(\lambda \kappa)}. 
                                                 \eqno{(3.27)} 
$$
It is convenient to write $ \delta {\cal M}_{Q}^{(1)}$ as 
$$
   \delta {\cal M}_{Q}^{(1)} = (A\mu^{2} + BQ^{2})
            {\rm log}\left({Q \over \Lambda}\right) +\ldots
                                                 \eqno{(3.28)} 
$$
where we omit terms that are constant or small. The coefficients $A$ 
and $B$ are given by
$$
A= { -3i\over 4\pi^{2}}[ (N+8){\lambda^{E}}^{2}
                            + 2(2N +1)\lambda ^{E} \kappa \mu^{2}]
                                                 \eqno{(3.29)} 
$$
and 
$$
B= { -3i\over 4\pi^{2}}(2N+7) \lambda ^{E} \kappa 
                                              Q^{2}.
                                                 \eqno{(3.30)} 
$$

The counterterms are determined at $Q^{2}=-\mu^{2}$ (or, more 
precisely, in a small neighborhood around $Q^{2}=-\mu^{2}$) from 
equation (3.4), which implies 
$$
-6i(\delta \lambda^{E} - \delta \kappa Q^{2})|_{Q^{2}=-\mu^{2}} =
            - \delta {\cal M}_{Q^{2}=-\mu^{2}}^{(1)}.
                                                 \eqno{(3.31)} 
$$
Equating powers of $Q^{2}$ the counterterms are then 
$$
\delta \lambda^{E} = -{i\over 6} A\mu^{2} {\rm log}\left({\mu \over 
                  \Lambda}\right)       \eqno{(3.32)} 
$$
and
$$
\delta \kappa  = +{i\over 6} B {\rm log}\left({\mu \over 
                  \Lambda}\right)       \eqno{(3.33)} 
$$                  
Then from (3.3) and (3.26) the running couplings are
$$
 \lambda^{E}_{Q} = \lambda^{E} +{i\over 6} A\mu^{2} {\rm log} 
                 \left({Q \over \mu}\right)       \eqno{(3.34)} 
$$
and
$$
\kappa_{Q}= \kappa - {i\over 6} B{\rm log} 
                \left({Q \over \mu}\right).   \eqno{(3.35)} 
$$  
or explicitly 
$$
\lambda^{E}_{Q}=  \lambda^{E} 
          +{1\over 8\pi^{2}}{\rm log}\left ({Q\over \Lambda} \right)
          \left\{ (N+8){\lambda^{E}}^{2} +2(2N+1)\lambda^{E} \kappa 
          \mu^{2}\right \}
                                                 \eqno{(3.36)} 
$$
and 
$$
\kappa^{E}_{Q}=  \kappa^{E} +
          {1\over 8\pi^{2}}{\rm log}\left ({Q\over \Lambda} \right)
          ( 2N+7) \lambda^{E} \kappa. 
                                                 \eqno{(3.37)} 
$$
To one loop and to leading order in $\kappa$, the coupled RGE's are 
then 
$$
{d\lambda^{E}_{Q}\over d{\rm log}Q} = 
    {1\over 8\pi^{2}}
    \left\{ (N+8){\lambda^{E}_{Q}}^{2} +2(2N+1)\lambda^{E}_{Q} \kappa _{Q} 
          \mu^{2}\right \}
                                                 \eqno{(3.38)} 
$$
and 
$$
{d\kappa^{E}_{Q}\over d{\rm log}Q} =
          {1\over 8\pi^{2}}
          \left( 2N+7\right) \lambda_{Q}^{E} \kappa_{Q}. 
                                                 \eqno{(3.39)} 
$$

\noindent {\bf (4) A Renormalizable Model }

In this section we construct a renormalizable model with a heavy 
scalar of mass $M\gg \mu$, which replicates the effective theory 
defined in equation (3.1) at low energy, $\mu < Q \ll M$, and then use 
the renormalizable model to verify the RGE's obtained in the previous 
section.  In the renormalizable model the one loop, log($Q$) dependent 
terms arise both from the low energy limit of finite Feynman diagrams 
as well as from log divergent diagrams reflecting the divergences of 
the original $\phi^{4}$ theory.  The results are of course regulator 
independent.  For convenience we use dimensional regularization here.  
The renormalizable model provides a very useful check on the 
calculation because the result arises in a rather different way from 
the effective theory --- in particular, the order $\mu^{2}/Q^{2}$ 
terms from equations (3.22 - 3.23) contribute differently, so that the 
results disagree if those terms are overlooked (as the author learned 
the hard way).

In addition to the O$(N)$ vector field $\phi$ we now add an O$(N)$ 
singlet scalar field $\sigma$.  The relevant terms in the 
renormalizable Lagrangian are
$$
{\cal L}_{R} = {1 \over 2}\left( (\partial {\phi})^{2} 
                                    +(\partial {\sigma})^{2}\right)
          - {\lambda^{R} \over 4} ({\phi}^{2})^{2}
           -{G\over 2}\sigma \phi^{2} 
          - {\mu^{2} \over 2}{\phi}^{2}
          -{M^{2} \over 2}{\sigma}^{2},    \eqno{(4.1)} 
$$
where $M \gg \mu$ and $G$ is a coupling constant with the dimension of
a mass.  Interaction terms involving more than a single heavy field
$\sigma$, such as $\sigma^{2}\phi^{2}$, $\sigma^{3}$ or $\sigma^{4}$
are neglected since they give rise to diagrams that are suppressed by 
additional powers of $M^{2}$, inducing corrections beyond the leading
order in $\kappa$ in the effective low energy theory.  We will see
that even in tree approximation the coupling $\lambda^{R}$ is not
equal to the analogous coupling $\lambda^{E}$ of the effective theory.

Despite our misleading notation, the renormalizable model is not a
``sigma model'' since in general the interactions do not have an
$O(N+1)$ symmetry. In order for ${\cal L}_{R}$ to be embedded in a
sigma model (i.e., one with its symmetry softly broken by the explicit
``pion'' mass $\mu$) the parameters would have to be related by
$\lambda^{R}=G^{2}/2M^{2}$. This is not a relevant limit for us since
it implies $\lambda^{E}=0$ (see equation (4.4) below), as required by
the low energy theorem for ``$\pi \pi$'' scattering.

In tree approximation the $ \phi_{1} \phi _{1} \ra \phi_{1} \phi _{1}$ 
scattering amplitude is
$$
{\cal M}(\phi_{1} \phi _{1} \ra \phi_{1} \phi _{1}) =
       - 6i\lambda^{R}  -iG^{2}\left(
       {1\over s - M^{2}} +  {1\over t - M^{2}} + {1\over u - M^{2}} 
       \right)           
                                  \eqno{(4.2)} 
$$   
Expanding for $M^{2} >> |s|,|t|,|u|$ this is 
$$
{\cal M}(\phi_{1} \phi _{1} \ra \phi_{1} \phi _{1} ) =- 6i\lambda^{R}  
           + 3i{G^{2} \over M^{2}} +i{G^{2}\over M^{4}}(s+t+u)
                                  \eqno{(4.3)} 
$$                            
which is equivalent to the tree approximation amplitude of the 
effective theory, equation (3.2), if we identify  
$$
\lambda^{E}= \lambda^{R} - {G^{2} \over 2M^{2}}
                                  \eqno{(4.4)} 
$$
and
$$
\kappa= {G^{2} \over 2M^{4}}.
                                  \eqno{(4.5)} 
$$
The first term in the expansion of the $\sigma$ propagator induces a 
tree level shift in $\lambda$ while the second term reproduces the 
dimension 6 operator in ${\cal L}_{\rm EFF}$.

We now consider the one loop corrections to the $\phi_{1} \phi _{1}
\ra \phi_{1} \phi _{1}$ scattering amplitude in the renormalizable
model.  Each of the six Feynman diagrams in the effective theory,
shown in figure 2b, is replaced by three diagrams in the toy model.
The first of these, shown in figure 4a, contains precisely the same
logarithmically divergent integral that renormalizes lambda in the
original $\phi^{4}$ theory, shown in figure 2a. In figure 4a it
corresponds to a $\sigma \phi \phi$ vertex correction.
Using the ${\overline {MS}}$ prescription and the symmetric, off-shell
external momenta defined in equation (3.3), the six diagrams of type
figure 4a together contribute
$$
{\cal M}_{a}= {-3i\lambda^{R}G^{2} \over 8\pi^{2}(Q^{2}-M^{2})} 
                     (N+2) \int^{1}_{0} dx\ {\rm log}\left({X\over 
                      \mu_{D}^{2}}\right)            \eqno{(4.6)} 
$$
where $\mu_{D}$ is the regulator scale.  Using equation (3.22) and 
expanding for $M^{2}\gg Q^{2}\gg \mu^{2}$ this becomes
$$
{\cal M}_{a}= { 3i\lambda^{R}G^{2} \over 4\pi^{2} }  
                     (N+2){\rm log}\left({Q\over \mu_{D}} \right)
                     \left\{ {1\over M^{2}} -2{\mu^{2}\over M^{4}}
                     +{Q^{2}\over M^{4}} \right\}
                                 \eqno{(4.7)} 
$$

In addition to ${\cal M}_{a}$ there are 12 finite diagrams of the type 
shown in figure 4b which together contribute
$$
{\cal M}_{b}= 36 \lambda^{R}G^{2}\int_{k}  {1 \over 
    ((k-p_{1})^{2}-\mu^{2})(k^{2}-M^{2})((k+p_{2})^{2}-\mu^{2}) }
                          \eqno{(4.8)} 
$$
or, introducing Feynman parameters and symmetrizing, 
$$
{\cal M}_{b}= 72 \lambda^{R}G^{2}\int^{1}_{0}dx
                         \int^{1-x}_{0}dy \int_{k}  {1 \over (k^{2}-Y)^{3} }
                          \eqno{(4.9)} 
$$
where
$$
Y= yM^{2} + (1-y)\mu^{2} 
        + \left[ {3\over 4}(y^{2}-y) +x^{2} -x +xy \right]Q^{2} 
                          \eqno{(4.10)}        
$$
After the $d^{4}k$ integration the result is 
$$
{\cal M}_{b}= {-9i \lambda^{R}G^{2}\over 4\pi^{2}}\int^{1}_{0}dx
                         \int^{1-x}_{0}dy {1 \over Y}
                          \eqno{(4.11)}        
$$
Performing the $y$ integration and retaining only the terms 
proportional to log($Q$) for $M\gg Q$, we find 
$$
{\cal M}_{b}= {9i \lambda^{R}G^{2}\over 4\pi^{2}M^{2}}
                        \int^{1}_{0}dx  \left\{  \left[ 
                      1+{\mu^{2}\over M^{2}} +{3\over 4}{Q^{2}\over M^{2}} 
                           \right]  -x{Q^{2}\over M^{2}} \right\} 
                           {\rm log}\left({X\over M^{2}}\right)
                          \eqno{(4.12)}        
$$
where $X$ is defined in equation (3.8). The terms that are omitted in 
(4.12) are either constants that are absorbed in counterterms or are 
of higher order in small ratios. Applying equations (3.22 - 3.23) the 
final result for ${\cal M}_{b}$ is
$$
{\cal M}_{b}= {3i \lambda^{R}G^{2}\over 4\pi^{2}M^{2}}
                       {\rm log}\left({Q\over M}\right)
                       \left\{ 6 +3{\mu^{2}\over M^{2}} 
                       +{3\over 2}{Q^{2}\over M^{2}} \right\}
                          \eqno{(4.13)}        
$$

The complete one loop amplitude is given by combining (4.7) and 
(4.13),
$$
\delta {\cal M}_{R}^{(1)}= {3i \lambda^{R}G^{2}\over 4\pi^{2}M^{2}}
                       {\rm log}\left({Q\over \mu}\right)
                       \left\{ (N+8) -(2N+1){\mu^{2}\over M^{2}} 
                          +\left(N+{7\over 2}\right){Q^{2}\over M^{2}} \right\}
                          \eqno{(4.14)}        
$$
where again we drop terms that do not vary as log($Q$) and will be 
absorbed in counterterms. 

The term proportional to (N+8) in (4.14) provides an interesting 
consistency check.  Since it is proportional to $G^{2}/M^{2} \propto 
\kappa M^{2}$ it seems to imply a nondecoupling contribution to the 
low energy theory which would invalidate the effective Lagrangian of 
section 3.  But it actually provides just the appropriate 
renormalization of the tree level shift encountered in equation (4.4),
$$
\delta \lambda_{\rm Tree}= -{G^{2}\over 2M^{2}},   \eqno{(4.15)}
$$
to give $\lambda^{E}_{Q}= \lambda^{R}_{Q} + 
\delta \lambda_{\rm Tree}$ the correct order ${\lambda^{E}}^{2}$ 
renormalization. That is, in the effective theory the order 
${\lambda^{E}}^{2}$ renormalization, from (3.30), is\footnote{
The shift in the scale of the logarithm from $\Lambda$ to $\mu$ is 
absorbed by the counterterm in the renormalization procedure.} 
$$
\delta \lambda^{E}_{Q}=  
          {N+8\over 8\pi^{2}}{\lambda^{E}}^{2}
          {\rm log}\left ({Q\over \mu} \right).   \eqno{(4.16)} 
$$
Substituting 
$$
{\lambda^{E}}^{2}\simeq{\lambda^{R}}^{2} + 
                2\lambda^{R}\delta \lambda_{\rm Tree}   \eqno{(4.17)}
$$
we see, using (4.15), that the contribution of the second term in 
(4.17) to (4.16) is 
$$
-{N+8\over 8\pi^{2}}{\lambda^{R} G^{2}\over M^{2}}
     {\rm log}\left ({Q\over \mu} \right). \eqno{(4.18)}
$$
But this is precisely equal to $i/6$ times the first term in (4.14), 
as required for consistency with equation (2.10).

The remaining two terms in (4.14), proportional to
$G^{2}\mu^{2}/M^{4}$ and $G^{2}Q^{2}/M^{4}$, agree precisely with the
corresponding terms proportional to $\kappa \mu^{2}$ and $\kappa
Q^{2}$ in the effective theory, as may be seen by comparing (4.14)
with equations (3.14) and (3.24) using the expression for $\kappa$ in
(4.5).  Therefore the renormalization of $\lambda^{E}$ and $\kappa$
computed in the effective theory in section 3 is verified by the
results obtained here from the renormalizable model.  We see that
renormalization effects which arise in the effective theory from a
combination of quadratically and logarithmically divergent diagrams
arise in the renormalizable model in a rather different way, with some
log($Q$) terms arising from diagrams with the usual ultraviolet
logarithmic divergences of the original $\phi^{4}$ theory (figure 4a)
while others arise from the low energy limit of finite diagrams
(figure 4b).

\noindent {\bf (5) Solution of the RGE's }

We now solve the coupled RGE's of the effective theory, first 
rewriting the RGE's, equations (3.32 - 3.33), in a more compact 
notation with the superscript $E$ suppressed,
$$
\lambda_{Q}^{\prime}= 
                          b_{N}\left( \lambda_{Q}^{2} + 
                          {2(N+2)\over N+8}\lambda _{Q} 
                          \gamma _{Q} \right)             \eqno{(5.1)}
$$ 
and 
$$
\gamma_{Q}^{\prime}= {2N+7 \over 8\pi^{2}}\lambda _{Q} \gamma _{Q} 
                                       \eqno{(5.2)}
$$ 
where $b_{N}$ was given in equation (2.13), and we have defined
$$
\gamma _{Q} = \kappa _{Q} \mu^{2}.            \eqno{(5.3)}
$$ 
The quantity $\gamma_{Q}$ has the same log($Q$) dependence as 
$\kappa_{Q}$ since $\mu$ has no log($Q$) dependence at one loop order.

We solve the coupled equations by constructing a quantity $f_{Q}$ that 
obeys the usual one loop RGE,
$$
f _{Q} = \lambda _{Q} + c_{N}\gamma _{Q},            \eqno{(5.4)}
$$ 
where $c_{N}$ is determined by requiring 
$$
f_{Q}^{\prime} = b _{N} f_{Q}^{2}.           \eqno{(5.5)}
$$ 
Working to first order in $\gamma_{Q}$ we find 
$$
c_{N}= {2\over 9}(2N+1).           \eqno{(5.6)}
$$ 
Integrating (5.5) from $-\mu^{2}$ to $Q^{2}$ we have the familiar 
solution
$$
f_{Q}= {f_{\mu} \over 1-b_{N}f_{\mu}
                  {\rm log}\left({Q \over \mu}\right) }.
           \eqno{(5.7)}
$$

To solve for $\lambda_{Q}$ and $\gamma_{Q}$ we replace $\lambda_{Q}$
in (5.2) by $f_{Q}$, valid to first order in $\gamma_{Q}$, obtaining
$$
\gamma_{Q}^{\prime}= {2N+7 \over 8\pi^{2}}f _{Q} \gamma _{Q}.  
                                       \eqno{(5.8)}
$$
Substituting (5.7) into (5.8) and integrating we then find
$$
\gamma_{Q}= \gamma_{\mu}\left[ {1 
                     \over 1-b_{N}f_{\mu}{\rm log}\left({Q \over \mu}\right) }
                           \right]^{2N+7\over N+8}
                                       \eqno{(5.9)}
$$ 
and, from (5.4), (5.6), and (5.9),
$$
\lambda_{Q}= {\lambda_{\mu} \over 
                 1-b_{N}f_{\mu}{\rm log}\left({Q \over \mu}\right) }
                 +{c_{N}\gamma_{\mu} \over 
                         1-b_{N}f_{\mu}{\rm log}\left({Q \over \mu}\right) }
                 \left\{1 - \left[ {1\over 
                 1-b_N f_{\mu}{\rm log}\left({Q \over \mu}\right) }
                         \right]^{N-1 \over N+8}   \right\}.  
                                        \eqno{(5.10)}
$$

We consider two special cases.  For $N=1$ the solutions take a 
particularily simple form,
$$
\gamma_{Q}|_{N=1}= {\gamma_{\mu} 
           \over 1-b_{1}f_{\mu}{\rm log}\left({Q \over \mu}\right) },
                                       \eqno{(5.11)}
$$
and 
$$
\lambda_{Q}|_{N=1}= {\lambda_{\mu} \over 
                 1-b_{1}f_{\mu}{\rm log}\left({Q \over \mu}\right) }.  
                                        \eqno{(5.12)}
$$
We also consider $N=4$ which would correspond to the 
Higgs sector of the Standard Model if we were to consider the broken 
symmetry phase.  Then $c_{4}=2$ and the running couplings are
$$
\gamma_{Q}|_{N=4}= \gamma_{\mu}\left[ {1 
                     \over 1-b_{4}f_{\mu}{\rm log}\left({Q \over \mu}\right) }
                           \right]^{5\over 4}.
                                       \eqno{(5.13)}
$$ 
and
$$
\lambda_{Q}|_{N=4}= {\lambda_{\mu} \over 
                 1-b_{4}f_{\mu}{\rm log}\left({Q \over \mu}\right) }
                 +{2\gamma_{\mu} \over 
                         1-b_{4}f_{\mu}{\rm log}\left({Q \over \mu}\right) }
                 \left\{1 - \left[ {1\over 
                 1-b_{4}f_{\mu}{\rm log}\left({Q \over \mu}\right) }
                         \right]^{1 \over 4}   \right\}.  
                                        \eqno{(5.14)}
$$

From these solutions to the RGE's we see that ``new physics'' 
represented by the dimension 6 operator has two effects on the Landau 
pole.  First, as we will discuss in the next section, it changes the 
relationship between the coupling constant $\lambda$ and the position 
of the pole.  Second it also affects the strength of the singularity 
at the pole, making the leading singularity stronger for all $N>1$.  
From (5.10) we see that since $c_{N}$ is proportional to $N$, the 
correction proportional to $c_{N}\gamma_{\mu}$ would dominate 
$\lambda_{Q}$ for sufficiently large $N$.  Our perturbative 
approximation then breaks down in the large $N$ limit, which would 
require a separate analysis.

\noindent {\bf (6) The Landau pole and the low energy coupling constant }

We now consider the effect of the dimension 6 operator on the 
relationship between the low energy coupling constant and the location 
of the Landau pole.  Without the dimension 6 operator, i.e., for 
$\kappa=0$, we see from equation (2.15) that the pole location is 
fixed by the low energy coupling constant and mass, 
$$
\Lambda_{\rm Landau}= 
              \mu\ {\rm exp}\left({1 \over b_{N}\lambda_{\mu}}
                                          \right).         \eqno{(6.1)}
$$
Equivalently, the low energy coupling is determined by the 
ratio $\Lambda_{\rm Landau}/ \mu$, 
$$
\lambda_{\mu}= 
             {1 \over b_{N}{\rm log}\left(
             { \Lambda_{\rm Landau} \over \mu}
                                          \right)}.         \eqno{(6.2)}
$$
New physics must intervene at or below the pole. Defining $M$ to 
be the new physics mass scale, (6.1) implies 
$$
M \leq \mu\ {\rm exp}\left({1 \over b_{N}\lambda_{\mu}}
                                          \right).         \eqno{(6.3)}
$$
In order that the low energy theory have some domain of validity we 
require a minimal hierarchy $R$ between $M$ and the mass 
scale of the low energy theory,
$$
{M \over \mu} \geq R,    \eqno{(6.4)}
$$
which implies the upper limit on the low energy coupling constant,
$$
\lambda_{\mu} \leq {1 \over b_{N}{\rm log}R }.    \eqno{(6.5)}
$$                                

These are simple but powerful relations. They are accurate in the 
perturbative domain, for small coupling $\lambda_{\mu}$ and large 
hierarchy $R$. At strong coupling and small $R$, the domain of the 
triviality bound, lattice simulations have found them to be 
qualitatively correct and, beyond that, accurate to about $\simeq 
30\%$.\cite{lattice}

The dimension 6 operator considered in section 3 modifies these 
relations, due to the effect of the high scale physics on the running 
of the scalar coupling constant. The result is simply to replace 
$\lambda_{\mu}$ in the above equations with $f_{\mu}$ defined in 
(5.5).  Then (6.1) and (6.2) become 
$$
\Lambda_{\rm Landau}= 
              \mu\ {\rm exp}\left({1 \over b_{N}
              (\lambda_{\mu}+c_{N}\kappa_{\mu}\mu^{2})}
                                            \right).        \eqno{(6.6)}
$$
and 
$$
\lambda_{\mu}= 
             {1 \over b_{N}{\rm log}\left(
             { \Lambda_{\rm Landau} \over \mu}
                                          \right)} 
              - c_{N}\kappa_{\mu}\mu^{2}.        \eqno{(6.7)}
$$
The upper bound on the low energy coupling constant becomes 
$$
\lambda_{\mu} \leq {1 \over b_{N}{\rm log}R }
           - c_{N}\kappa_{\mu}\mu^{2}.    \eqno{(6.8)}
$$

The sign of the new physics correction depends then on the sign of 
$\kappa_{\mu}$.  If $\kappa_{\mu} > 0$ the Landau pole position is 
lowered for fixed $\lambda_{\mu}$ and $\mu$, and the upper bound on 
$\lambda_{\mu}$ becomes stronger for given hierarchy $R$.  Conversely 
for $\kappa_{\mu} < 0$ the Landau pole moves to higher energy and the 
upper bound on $\lambda_{\mu}$ is weakened.  In the renormalizable 
model considered in section 4, in which the new physics arises from 
the exchange of a heavy singlet scalar, equation (4.5) implies 
$\kappa_{\mu}>0$.  In general the sign may be positive or negative.

If the dimension 6 operator arises from a dimension 4 interaction with
dimensionless coupling $g$ between high-scale quanta of mass $M$
and the light scalar fields $\phi$, then $\kappa$ will be of order
${\rm O} (g^{2}/M^{2})$ where for strong coupling $g^{2}$ would be of
order O($4\pi$).  If the underlying interaction has dimension 3 with
dimensionful coupling $G$, then $\kappa \simeq {\rm O}(G^{2}/M^{4})$,
as in equation (4.5).  For strong coupling we would then expect
$G^{2}/M^{2}\simeq {\rm O}(4\pi)$, and, again, $\kappa \simeq {\rm
O}(4\pi/M^{2})$.

In general the strength of the dimension 6 operator is characterized 
by a dimensionless quantity $C$, defined by
$$
\kappa_{\mu}= {C \over M^{2}}.    \eqno{(6.9)}
$$
Assuming now that $M$ is as heavy as it can be, $M\simeq \Lambda_{\rm
 Landau}$, and that the hierarchy inequality (6.4) is also saturated,
i.e.,
$$
{M^2 \over \mu^2} = { \Lambda_{\rm Landau}^2 \over \mu^2} = R
                          \eqno{(6.10)},
$$   
then (6.7) becomes 
$$
\lambda_{\mu} = {1 \over b_{N}{\rm log}R }
           - c_{N}{C \over R^{2}}.    \eqno{(6.11)}
$$
Defining $r_{\lambda}$ as the ratio of the value of 
$\lambda_{\mu}$ determined from (6.11) to the corresponding value,
equation (6.2), for $C=\kappa_{\mu}=0$, we have
$$
r_{\lambda}= 1 - b_{N}c_{N}C{ {\rm log} R \over R^{2}}.    
                                                    \eqno{(6.12)}
$$
Similarly, we define $r_{\Lambda}$ as the ratio of  
$\Lambda_{\rm Landau}$ determined for $\kappa_{\mu} \neq 0$ from (6.6) 
relative to the value for $\kappa_{\mu}=0$ from (6.1), for the same 
values of $\lambda_{\mu}$ and $\mu$, and find  
$$
r_{\Lambda}= {\rm exp}\left( -{c_{N}C\over b_{N}R^{2}}
              {1 \over \lambda_{\mu}( \lambda_{\mu}+c_{N}{C\over 
                        R^{2}} )} \right).
                                                    \eqno{(6.13)}
$$
Expanding to leading order in $\kappa_{\mu} \mu^{2}/\lambda_{\mu}$, 
$r_{\Lambda}$ is approximately
$$
r_{\Lambda}\simeq 1 - b_{N}c_{N}C \left({ {\rm log} R \over 
R}\right)^{2}, \eqno{(6.14)}
$$
which is enhanced by an additional factor of ${\rm log} R$ relative to 
the correction to $r_{\lambda}$ in (6.12).

The value of $\lambda_{\mu}$ from equation (6.11) for $N=4$ is
plotted in figure 5 for $C = 0, \pm 4\pi$ as a function of the
hierarchy $R$. Numerical values for the ratios $r_{\lambda}$ and
$r_{\Lambda}$ from equations (6.12) and (6.13) are given in tables 1
and 2 for $N=4$ and $N=1$ respectively with $C=\pm 4\pi$.  For $N=1$
the corrections are approximately four times smaller than for $N=4$.
For $N=4$ the corrections to the coupling constant are
sizeable, reaching 66\% for the hierarchy $R=2$ at which the
triviality bound is customarily obtained in lattice calculations.  The
value of such a large correction cannot be taken literally since it
exceeds the domain of validity of the perturbative
approximation, but it suggests that large corrections, potentially
even of order one, are possible.  At larger values of $R$ the
corrections become smaller and are therefore more reliably known.  We
may for instance consider where the approximation (6.14) becomes a
good description of (6.13).  From table 1 we find that this occurs at
$R=5$, where (6.14) implies a 40\% correction to $r_{\Lambda}$ in good
agreement with the result shown in the table, and for which the
correction to the triviality bound is 25\%.

\noindent {\bf (7) Conclusion}

We have studied the effect of new physics on the RG analysis of the 
Landau pole and the triviality bound in the unbroken phase of O($N$) 
$\phi^{4}$ theory.  Including a dimension 6 operator to represent 
the low energy effects of the new physics that must exist at the 
Landau pole, we find that the pole position and the upper bound on the 
coupling constant can be modified by substantial amounts, if the new 
physics is strongly coupled to the O($N$) scalars and if the O($N$) 
scalars are themselves strongly coupled.  The analysis is performed in 
the spirit of the original Dashen-Neuberger analysis, to explore the 
possible order of magnitude of the effects in a simple 
approximation.

Quantitative results for the strong coupling regime would require 
lattice simulations.  As discussed in the introduction, the related 
lattice simulations carried out in \cite{hknv} had a different goal 
than ours --- to see the effect of dimension 6 operators subject to 
the constraint that the low energy Higgs sector resemble the SM Higgs 
sector to within a few percent --- and is not directly comparable to 
the calculation presented here since different phases are considered 
and the required operator matching has not been done.  Here, within 
the limitations of one loop perturbation theory, we explored the 
maximum effect on the bound without regard to the size of other 
corrections to the low energy physics.  It would be interesting to 
study this regime with lattice simulations.  

To check the RG analysis of the effective theory we also considered a 
simple renormalizable model of the new physics, consisting of a heavy 
O($N$) singlet scalar field which in its low energy limit gives rise 
to the dimension 6 operator.  The model provides a computational check 
and a measure of physical insight.  We saw that in addition to giving 
rise to the dimension 6 operator, the exchange of the heavy scalar 
causes a tree-level shift $\delta \lambda_{\rm Tree}$ in the 
$\phi^{4}$ coupling constant $\lambda$ and that apparently dangerous 
one loop corrections, proportional to $\kappa M^{2}$ (where $M$ is the 
mass of the heavy scalar) conspire to give $\delta \lambda_{\rm Tree}$  
precisely the usual O($\lambda^{2}$) renormalization.

The sign of $\kappa$ determines whether the triviality limit is 
increased or decreased by new physics.  For the model considered here 
with a heavy O($N$) singlet scalar, $\kappa$ is positive, in which 
case the Landau pole position and the triviality bound on the coupling 
are both lowered.  For negative $\kappa$ the opposite occurs.  It is 
then very interesting to exhibit theories with $\kappa <0$ or to prove 
that none exist.  The lattice studies of dimension 6 operators in the 
Higgs phase reported possible increases of the triviality 
bound\cite{hknv}, but they too introduced dimension 6 operators by 
hand and so also did not address this issue. Since $\kappa$ is a 
parameter of an effective theory with a limited domain of validity, 
which might furthermore arise from the low energy limit of another 
effective theory with a still limited (though higher energy) domain 
of validity, it is likely that no general theorem exists. 

In the renormalizable O($N$) $\phi^{4}$ scalar theory it is easy to 
see that the RG flow of the coupling constant is the same in the 
symmetric and Higgs phases of the theory.  Because the dimension 6 
$\phi^4$ interaction gives rise to dimension 5 $\phi^3$ interactions in 
the Higgs phase, it is not immediately apparent that the same is true 
of the effective theory.  This question is under study and will be 
reported elsewhere.

\vskip 0.2in

\noindent {\bf Acknowledgements} I wish to thank Korkut Bardakci, 
Sidney Coleman, Mary Gaillard, Lawrence Hall, and Hitoshi Murayama for 
helpful discussions.

\newpage
\vskip 0.5in

Table 1.  Tabulation of $r_{\lambda}$ and $r_{\Lambda}$, from 
equations (6.12 - 6.13), for the $N=4$ theory with $C= \pm 4\pi$.

\begin{center}
\vskip 20pt
\begin{tabular}{c|cc|cc}
	& $C\ \ =$ &$4\pi$
	& $C\ \ =$ &$-4\pi$  \\
\hline
\hline
{$R$} &{$r_{\lambda}$} &{$r_{\Lambda}$} 
&{$r_{\lambda}$} &{$r_{\Lambda}$}\\
\hline
\hline
2 & 0.34 & 0.26 & 1.66 & 1.32 \\
\hline
$e$ & 0.48 & 0.34 & 1.52 & 1.41 \\
\hline
5 & 0.75 & 0.59 & 1.25 & 1.37 \\
\hline
7 & 0.85 & 0.71 & 1.15 & 1.29 \\
\hline
10 & 0.91 & 0.80 & 1.09 & 1.20 \\
\hline
\hline 
\end{tabular}
\end{center}

\vskip 2 in

Table 2.  Tabulation of $r_{\lambda}$ and $r_{\Lambda}$, from  
equations (6.12 - 6.13), for the $N=1$ theory with $C= \pm 4\pi$.

\begin{center}
\vskip 20pt
\begin{tabular}{c|cc|cc}
	& $C\ \ =$ &$4\pi$
	& $C\ \ =$ &$-4\pi$  \\
\hline
\hline
{$R$} &{$r_{\lambda}$} &{$r_{\Lambda}$} 
&{$r_{\lambda}$} &{$r_{\Lambda}$}\\
\hline
\hline
2 & 0.84 & 0.87 & 1.17 & 1.10\\
\hline
$e$ & 0.87 & 0.86 & 1.13 & 1.12 \\
\hline
5 & 0.94 & 0.90 & 1.06 & 1.10 \\
\hline
7 & 0.96 & 0.93 & 1.04 & 1.07 \\
\hline
10 & 0.98 & 0.95 & 1.02& 1.05\\
\hline
\hline 
\end{tabular}
\end{center}


\begin{figure}
\begin{center}
\includegraphics[height=2in,width=2in,angle=-90]{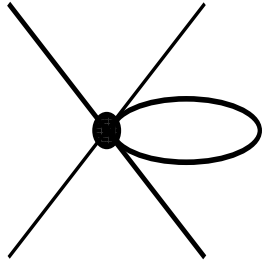}
\end{center}
\caption{One loop correction to the four point function 
from the operator $\phi^6/\Lambda^2$.}
\label{fig1}
\end{figure}

\begin{figure}
\begin{center}
\includegraphics[height=3in,width=3in,angle=-90]{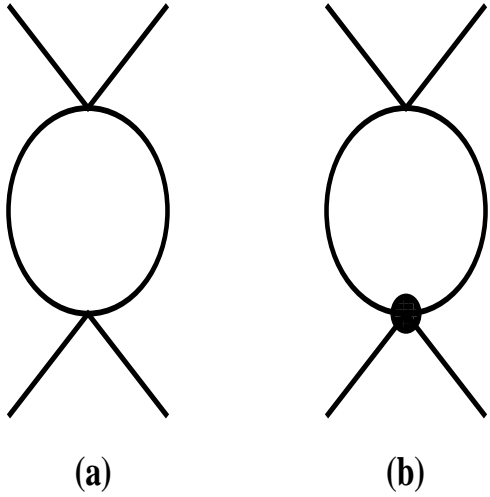}
\end{center}
\caption{Feynman diagrams for (a) the order $\lambda^2$ and 
(b) the order $\kappa \lambda$ corrections to the scattering 
amplitude. The black dot denotes the $(\partial \phi^2)^2$ interaction.}
\label{fig2}
\end{figure}

\begin{figure}
\begin{center}
\includegraphics[height=3in,width=2in,angle=-90]{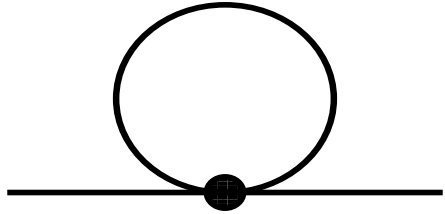}
\end{center}
\caption{The one loop correction to the self energy from
the $(\partial \phi^2)^2$ interaction.}
\label{fig3}
\end{figure}

\begin{figure}
\begin{center}
\includegraphics[height=4in,width=3in,angle=-90]{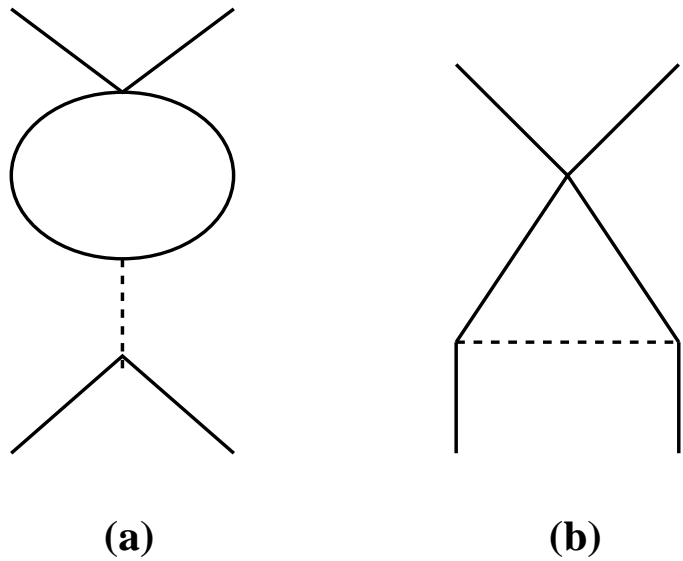}
\end{center}
\caption{The leading one loop corrections to $\phi \phi$
scattering from exchange of the heavy scalar $\sigma$,  
denoted by the dashed lines.}
\label{fig4}
\end{figure}

\begin{figure}
\begin{center}
\includegraphics[height=5.5in,width=4.5in,angle=90]{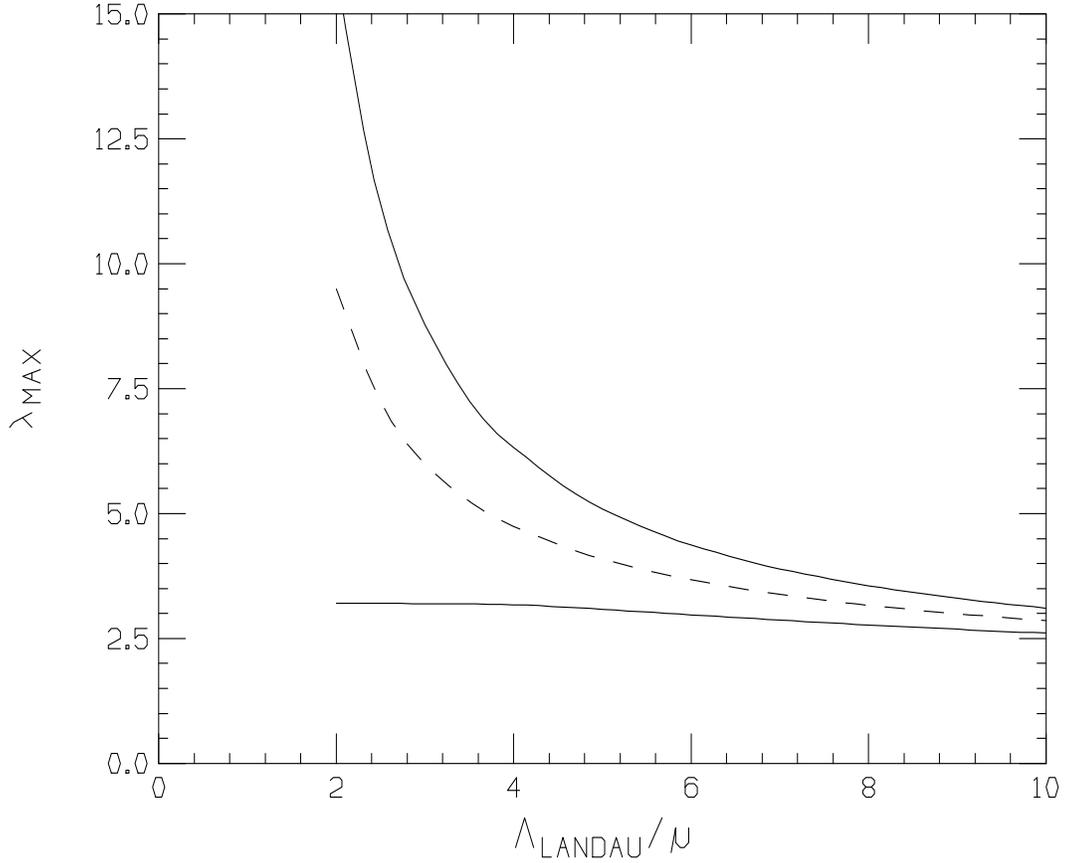}
\end{center}

\caption{The triviality upper limit on the coupling constant
$\lambda_{\mu}$ for $N=4$ as a function of the hierarchy $\Lambda_{\rm
Landau}/\mu$.  The dashed line shows the limit in the absence of the
dimension 6 operator, while the upper and lower solid lines are from
equation (6.11) with $C=-4\pi$ and $C=+4\pi$ respectively.}

\label{fig5}
\end{figure}

\end{document}